\documentclass{emulateapj}
\usepackage{amsmath}
\usepackage{graphicx}
\usepackage{natbib}
\usepackage[scaled]{helvet}
\usepackage{epsfig}
\usepackage{url}
\bibpunct{(}{)}{;}{a}{}{,}
\newcommand{\kms}{\,km\,s$^{-1}$}
\usepackage{textcomp}
\usepackage{mathpazo}
\usepackage{xspace}

\usepackage{color}
\usepackage[normalem]{ulem}

\newcommand{\bjdtdb}{\ensuremath{\rm {BJD_{TDB}}}}
\newcommand{\feh}{\ensuremath{\left[{\rm Fe}/{\rm H}\right]}}
\newcommand{\mh}{\ensuremath{\left[{\rm m}/{\rm H}\right]}}
\newcommand{\teff}{\ensuremath{T_{\rm eff}}}
\newcommand{\logg}{\ensuremath{\log g}}

\newcommand{\msun}{\ensuremath{\,M_\Sun}}
\newcommand{\rsun}{\ensuremath{\,R_\Sun}}
\newcommand{\lsun}{\ensuremath{\,L_\Sun}}

\newcommand{\rj}{\ensuremath{\,R_{\rm J}}}
\newcommand{\rearth}{\ensuremath{\,R_{\rm \Earth}}\xspace}
\newcommand{\mearth}{\ensuremath{\,M_{\rm \Earth}}\xspace}

\newcommand{\Kepler}{{\it Kepler}}
\newcommand{\ms}{\,m\,s$^{-1}$}

\newcommand{\thisstar}{HD~106315\xspace}
\newcommand{\mstar}{\ensuremath{M_{*}}}
\newcommand{\ar}{\ensuremath{a/R_*}}

\begin{document}

\title{A Multi-Planet System Transiting the $V$ = 9 Rapidly Rotating F-Star HD 106315}
\author{Joseph E. Rodriguez$^1$, George Zhou$^{1}$, Andrew Vanderburg$^{1}$, Jason D. Eastman$^{1}$,Laura Kreidberg$^{1}$, Phillip A. Cargile$^{1}$, Allyson Bieryla$^{1}$, David W. Latham$^{1}$, Jonathan Irwin$^{1}$, Andrew W. Mayo$^{1}$, Michael L. Calkins$^{1}$, Gilbert A. Esquerdo$^{1}$, Jessica Mink$^{1}$}

\affil{$^{1}$Harvard-Smithsonian Center for Astrophysics, 60 Garden St, Cambridge, MA 02138, USA}

\shorttitle{The \thisstar\ Planetary System}
\shortauthors{Rodriguez et al.}

\begin{abstract}

We report the discovery of a multi-planet system orbiting \thisstar, a rapidly rotating mid F-type star, using data from the K2 mission. \thisstar\ hosts a $2.51\pm0.12\,R_\oplus$ sub-Neptune in a 9.5 day orbit, and a $4.31_{-0.27}^{+0.24}\,R_\oplus$ super-Neptune in a 21 day orbit. The projected rotational velocity of \thisstar (12.9 \kms) likely precludes precise measurements of the planets' masses, but could enable a measurement of the sky-projected spin-orbit obliquity for the outer planet via Doppler tomography. The eccentricities of both planets were constrained to be consistent with 0, following a global modeling of the system that includes a \textit{Gaia} distance and dynamical arguments. The \thisstar system is one of few multi-planet systems hosting a Neptune-sized planet for which orbital obliquity measurements are possible, making it an excellent test-case for formation mechanisms of warm-Neptunian systems. The brightness of the host star also makes \thisstar c a candidate for future transmission spectroscopic follow-up studies.   

\end{abstract}


\section{Introduction}
The discovery of ``hot Jupiters'', gas giant planets orbiting very close to their host star, completely changed our understanding of planet formation. While it was widely believed that Jupiter-mass planets could only form far from their host stars, the discovery of these short-period massive planets demonstrated that planet formation and migration is a far more complex and dynamic process than previously thought. Theories for the origins of hot Jupiters include the idea that they form farther out in the protoplanetary disk, and through various mechanisms, migrate inward. However, the large number of compact planetary systems with Neptune-sized objects discovered in the last decade have led to the idea that smaller gaseous planets in close-in orbits might form in situ \citep{Chianglaughlin:2013,Boley:2016}, and the notion that hot Jupiters might form this way as well \citep{Batygin:2016}. \citet{Lee:2014} showed that run-away accretion of hydrogen gas around rocky cores can occur in the inner disk (0.1 AU), leading to planets with significant gaseous envelopes. \citet{Huang:2016} found that giant planets outward of 10-day period orbits are preferentially found in multi-planet systems, as opposed to the population of typically-lonely hot-Jupiters and hot-Neptunes. This suggests that hot Jupiters might have migrated inward chaotically, disrupting any inner planetary systems, while warm Neptunes perhaps formed in situ, often within compact planetary systems.


Tracers for the formation mechanisms of these planetary systems lie in their present-day orbital configurations, so characterizable transiting systems are key for us to discern the origins of these warm-Neptune / super-Earth systems. \citet{hansen:2013} proposed that compact super-Earth systems that formed in situ should exhibit small non-zero eccentricity distributions. Measurements of the angle between the orbital angular momentum vector and the host star's rotational axis ---its obliquity angle--- is another major observational clue to the way planets form and migrate. Lonely planets in high obliquity orbits could have been dynamically injected inward by outer companions \citep[e.g. HAT-P-11b][]{Hirano:2011,Sanchis-Ojeda:2011} while co-planar, high obliquity planetary systems \citep[e.g. Kepler-56, ][]{Huber:2013} may have formed within a torqued disk that was excited through interactions with stellar companions or magnetic fields \citep{batygin:2012, Spalding:2014}. Such torquing and warping of circumstellar disks around young stars has been well documented with high-resolution imaging by the Hubble Space Telescope (Beta Pictoris, \citealp{Kalas:1995, Heap:2000}) and by the millimeter mapping with the Atacama Large Millimeter/submillimeter Array (ALMA) (HD 142527, \citealp{Marino:2015}).  

\begin{figure*}[!ht]
\vspace{0.3in}
\includegraphics[width=0.99\linewidth, trim = 0 0.0in 0 0]{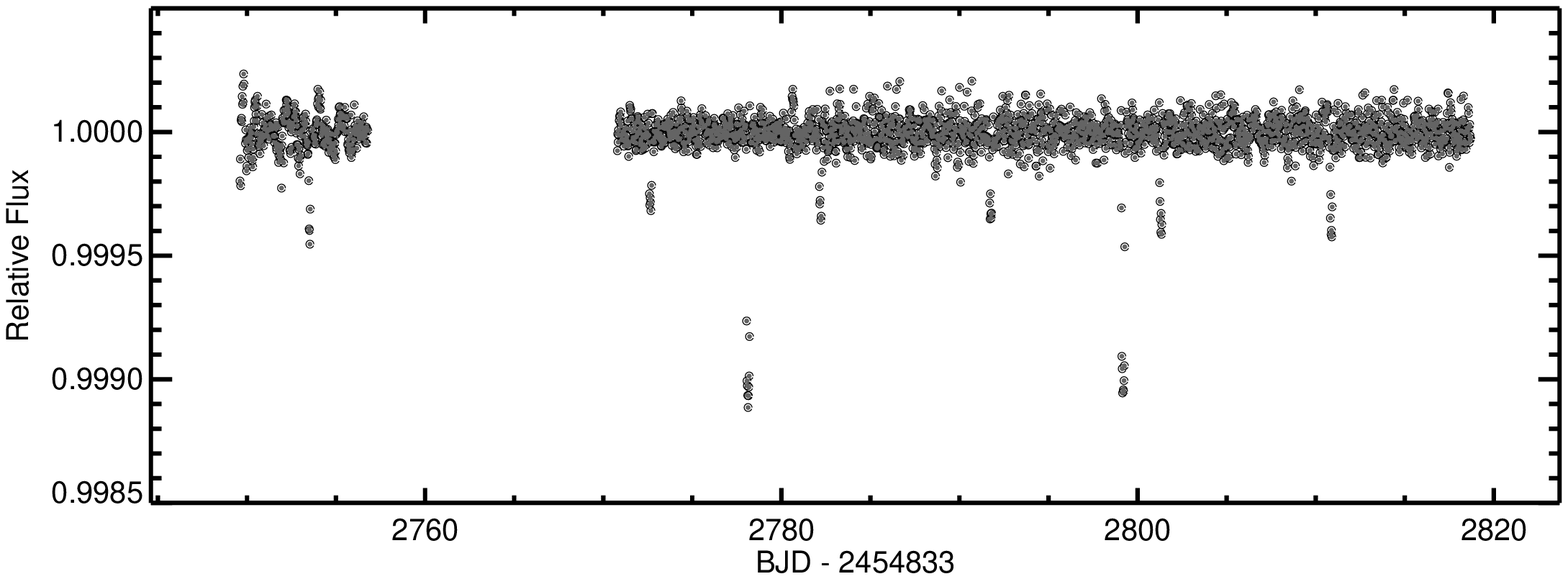}
\includegraphics[width=0.5\linewidth]{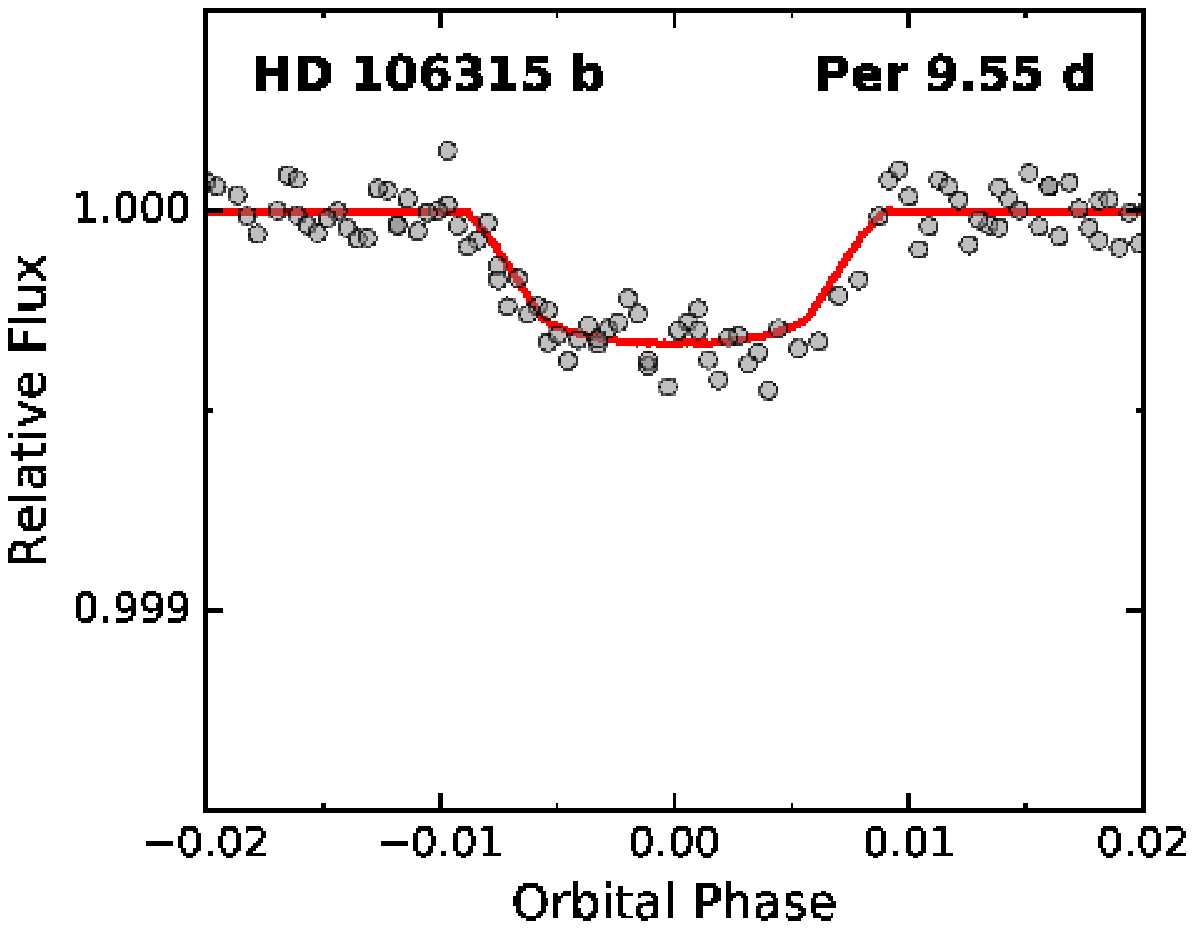} 
\includegraphics[width=0.5\linewidth]{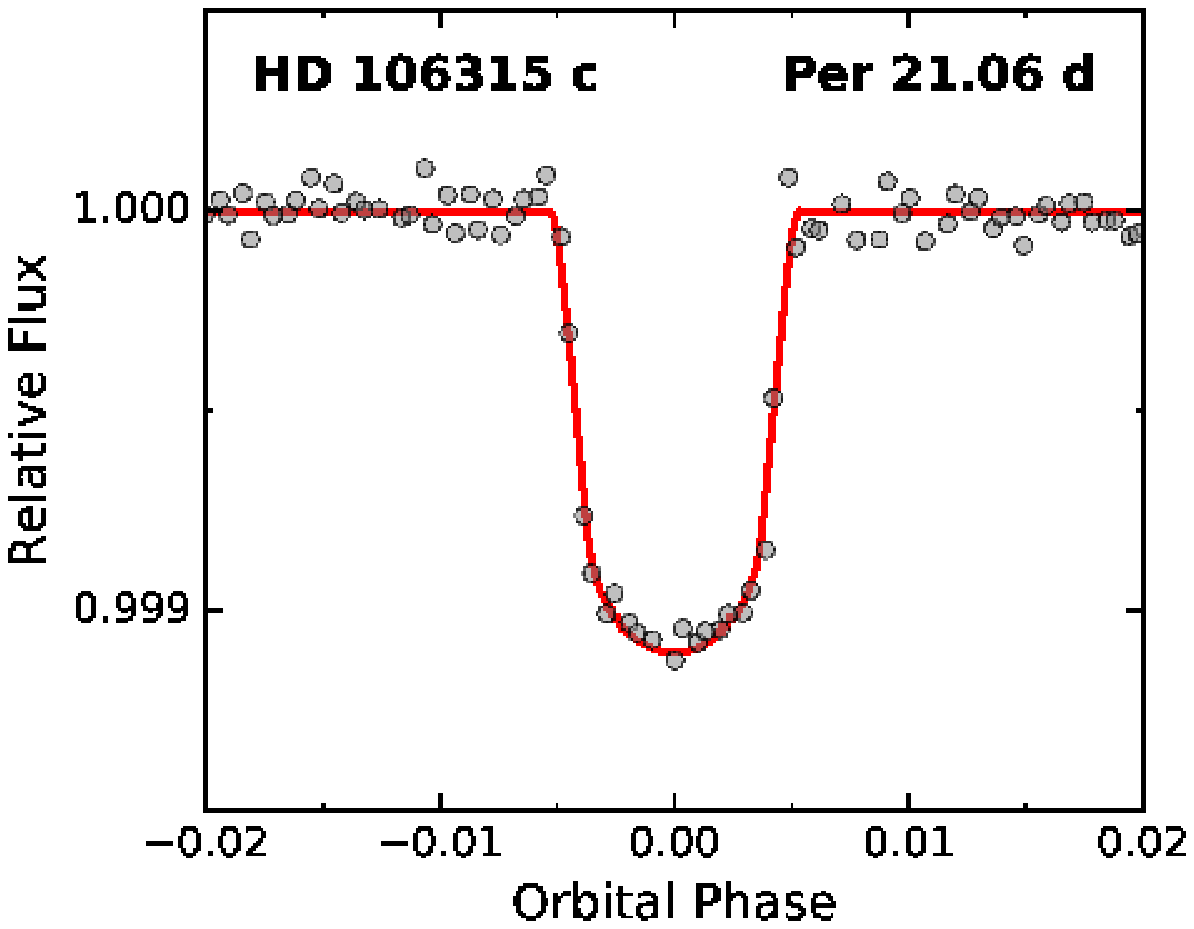}
\caption{(Top) The corrected K2 lightcurve for HD 106315 using the technique described in \citet{Vanderburg:2014}. (Bottom) Phase folded \emph{K2} light curves of HD 106315 b and c. The observations are plotted in grey, and the best fit model is plotted in red.}
\label{figure:LC}
\end{figure*}

Recently, measurements of the Rossiter-McLaughlin effect \citep{Rossiter:1924,McLaughlin:1924} and Doppler tomography \citep{Collier:2010} for transiting hot Jupiters have revealed a pattern that planet orbits and host star spins tend to be well aligned for cool stars, and misaligned for hot stars \citep{Winn:2010b,Albrecht:2012}. The change from aligned systems to misaligned systems appears to happen at a \teff $\sim$ 6250 K, where there is a known transition between slowly and rapidly rotating stars referred to as the Kraft break \citep{Kraft:1967, Kraft:1970}. It is believed that the lack of a convection zone in rapidly rotating stars above the Kraft break allows them to conserve more angular momentum from formation and have weaker magnetic fields resulting in less magnetic braking from stellar wind \citep{vanSaders:2013}. Similarly, observations of young disk-bearing T Tauri and Herbig Ae/Be stars have shown that the magnetic field strength decreases with stellar mass \citep{Gregory:2012, Alecian:2013}. \citet{Albrecht:2012} suggest that this trend in spin/orbit misalignments could be the signature of planet-planet scattering bringing hot Jupiters close into their host stars on highly-eccentric orbits, where the stronger tidal interactions with stars having a convective zone below the Kraft break bring the envelopes of cool stars back into alignment with the planetary orbits. However, \citet{Mazeh:2015} and \citet{Li:2016} used photometric spot modulation of \emph{Kepler} systems to show that this trend continues for smaller-planets in longer-period orbits, where tidal and magnetic interactions are significantly weaker, and realignment of the stellar envelope is unlikely. Precise spectroscopic obliquity measurements of smaller radius or longer period planetary systems around hot stars are key to solving this problem.

Few planetary systems with small or long-period planets transiting hot stars are amenable to such detailed characterization to discern their origins. Ground-based transit surveys are mostly sensitive to Jupiter-sized objects with short orbital periods, and while NASA's \Kepler\ mission has found many systems of small and long-period planets, most of the host stars are too faint for follow-up observations. The K2 mission and the upcoming Transiting Exoplanet Survey Satellite (TESS) mission \citep{Ricker:2015} offer opportunities to find small planets transiting bright stars above/below the Kraft break by virtue of looking at many more bright stars than the \Kepler\ mission did. K2 has yielded over a hundred planet discoveries, many of which orbit bright stars \citep{Sinukoff:2016, Crossfield:2016, Vanderburg:2016c}, and TESS is expected to find over a thousand planets transiting the closest and brightest stars in the sky \citep{sullivan:2015}. By virtue of their high photometric precision and wide-field survey designs, these missions will discover small planets transiting hot stars amenable to spectroscopic measurements of obliquity. 

In this paper, we present the discovery of two transiting planets orbiting the bright, $V$ = 9 star \thisstar\ (Table \ref{tab:LitProps}) from the K2 mission. The close-in planet, \thisstar\ b is a 2.5$\pm$0.1 \rearth sub-Neptune on a $9.5539_{-0.0007}^{+0.0009}$ day period, while the outer planet, \thisstar\ c,  is a $4.3_{-0.3}^{+0.2}$ \rearth super-Neptune in a $21.058\pm0.002$ day period. The host star has a temperature above the Kraft break, and as such is rapidly rotating at a projected speed of 12.9$\pm$0.4 \kms. The rapid rotation, brightness of the host star, and depth of the outer planet's transit make Doppler Tomography observations to determine the spin-orbit misalignment of the \thisstar\ system possible with high-resolution spectrographs on moderate aperture telescopes. \thisstar\ c could be the first warm Neptune-sized planet orbiting a star above the Kraft break with a measured spin-orbit angle, providing crucial information to its formation and evolutionary history.


\begin{table}
\footnotesize
\centering
\caption{\thisstar\ Magnitudes and Kinematics}
\begin{tabular}{llcc}
  \hline
  \hline
Other identifiers\dotfill & 
        \multicolumn{3}{l}{\thisstar} \\
      & \multicolumn{3}{l}{TYC 4940-868-1}				\\
      & \multicolumn{3}{l}{EPIC 201437844}				\\
	  & \multicolumn{3}{l}{2MASS J12135339-0023365}
\\
\hline
Parameter & Description & Value & Ref. \\
\hline
$\alpha_{J2000}$\dotfill	&Right Ascension (RA)\dotfill & 12:13:53.394			& 1	\\
$\delta_{J2000}$\dotfill	&Declination (Dec)\dotfill & -00:23:36.54			& 1	\\
\\
\\
B$_T$\dotfill			&Tycho B$_T$ mag.\dotfill & 9.488 $\pm$ 0.022		& 1	\\
V$_T$\dotfill			&Tycho V$_T$ mag.\dotfill & 9.004 $\pm$ 0.018		& 1	\\
\\
J\dotfill			& 2MASS $J$ mag.\dotfill & 8.116  $\pm$ 0.026		& 2, 3	\\
H\dotfill			& 2MASS $H$ mag.\dotfill & 7.962 $\pm$ 0.040	& 2, 3	\\
K$_S$\dotfill			& 2MASS $K_S$ mag.\dotfill & 7.853 $\pm$ 0.020	& 2, 3	\\
\\
\textit{WISE1}\dotfill		& \textit{WISE1} mag.\dotfill & 7.805 $\pm$ 0.023		& 4, 5	\\
\textit{WISE2}\dotfill		& \textit{WISE2} mag.\dotfill & 7.857 $\pm$ 0.019		& 4, 5 \\
\textit{WISE3}\dotfill		& \textit{WISE3} mag.\dotfill &  7.856 $\pm$ 0.023		& 4, 5	\\
\textit{WISE4}\dotfill		& \textit{WISE4} mag.\dotfill & 7.898$\pm$0.246		& 4, 5	\\
\\
$\mu_{\alpha}$\dotfill		& Gaia DR1 proper motion\dotfill & -1.678 $\pm$ 0.636 		& 6 \\
                    & \hspace{3pt} in RA (mas yr$^{-1}$)	& & \\
$\mu_{\delta}$\dotfill		& Gaia DR1 proper motion\dotfill 	&  11.912 $\pm$ 0.460 &  6 \\
                    & \hspace{3pt} in DEC (mas yr$^{-1}$) & & \\
\\
$RV$\dotfill & Systemic radial \hspace{9pt}\dotfill  & -3.65$\pm$0.1 & \S\ref{sec:Spec} \\
     & \hspace{3pt} velocity (\kms)  & & \\
$v\sin{i_\star}$\dotfill &  Rotational velocity \hspace{9pt}\dotfill &  12.9$\pm$0.4 \kms & \S\ref{sec:Spec} \\
$\mh$\dotfill &   Metallicity \hspace{9pt}\dotfill & -0.27$\pm$0.08 & \S\ref{sec:Spec} \\
$\teff$\dotfill &  Effective Temperature \hspace{9pt}\dotfill &  6251$\pm$52 K & \S\ref{sec:Spec} \\
log(g)\dotfill &  Surface Gravity \hspace{9pt}\dotfill &  4.1$\pm$0.1 (cgs) & \S\ref{sec:Spec} \\
V$_{mac}$\dotfill &  Macroturbulent Velocity \hspace{9pt}\dotfill &  4.0$\pm$0.3 \kms & \S\ref{sec:Spec} \\
V$_{mic}$ (fixed)\dotfill &  Microturbulent Velocity \hspace{9pt}\dotfill &  1.9 \kms  & \S\ref{sec:Spec} \\
$d$\dotfill & Distance (pc)\dotfill & $107.3 \pm 3.9$ & 6 \\
Spec. Type\dotfill & Spectral Type\dotfill & F5V & 7 \\
\hline
\hline
\end{tabular}
\begin{flushleft} 
 \footnotesize{ \textbf{\textsc{NOTES:}}
    References are: $^1$\citet{Hog:2000}, $^2$\citet{Cutri:2003}, $^3$\citet{Skrutskie:2006}, $^4$\citet{Wright:2010}, $^5$\citet{Cutri:2014}, $^6$\citet{Brown:2016} Gaia DR1 http://gea.esac.esa.int/archive/ , $^7$\citet{Houk:1999}.
}
\end{flushleft}
\label{tab:LitProps}
\end{table}

\section{Observations}
\subsection{K2 Photometry}
After the failure of its second fine-pointing reaction wheel, the \Kepler\ spacecraft has been re-purposed to obtain highly precise photometric observations of a set of fields near the ecliptic for its extended K2 mission \citep{Howell:2014}. As part of K2 Campaign 10, the \Kepler\ telescope observed \thisstar\ between 6 July 2016 and 20 September 2016. Usually, K2 observes targets continuously for 80 days at a time, but during Campaign 10, the spacecraft suffered several unanticipated anomalies. During the first six days of the campaign (6 July 2016 to 13 July 2016) the the spacecraft's pointing was off by about 12 arcseconds, causing many targets to fall at least partially outside of their ``postage stamp'' apertures. The pointing was then corrected, and Kepler observed for about 7 more days until one of the spacecraft's CCD modules failed on 20 July 2016. The module failure caused \Kepler\ to go into safe mode, and data collection was halted until 3 August 2016, at which point data collection proceeded normally until the end of the campaign on 20 September 2016. Upon public release of the K2 campaign 10 dataset, we downloaded the target pixel files, produced light curves, corrected for \Kepler's unstable pointing precision and known systematics using the technique described in \citet{Vanderburg:2014}, and searched for transiting planets with the pipeline described by \citet{Vanderburg:2016}. Our transit search identified two transiting planet candidates orbiting \thisstar: a candidate super-Neptune in a 21 day orbit, and a candidate sub-Neptune in a 9.5 day orbit. We then re-processed the K2 light curve by removing 4-$\sigma$ upwards outliers and the two largest single-point downwards outliers, and simultaneously fitting the transit signals, K2 systematics, and long-term flux variations using the method described by \citet{Vanderburg:2016}. For our analysis, we only use the data collected after the pointing correction on 13 July 2016, a total of 2506 data points taken over the course of 69 days (See Figure \ref{figure:LC}). We flattened the light curves for modeling by dividing away the best-fit low-frequency variations (which we modeled as a basis spline with breakpoints every 0.75 days) from our simultaneously-fit reprocessed light curve.

\begin{figure}[!ht]
\vspace{0.3in}
\centering\includegraphics[width=0.99\linewidth, angle = 0, trim = 0 0 0 1in]{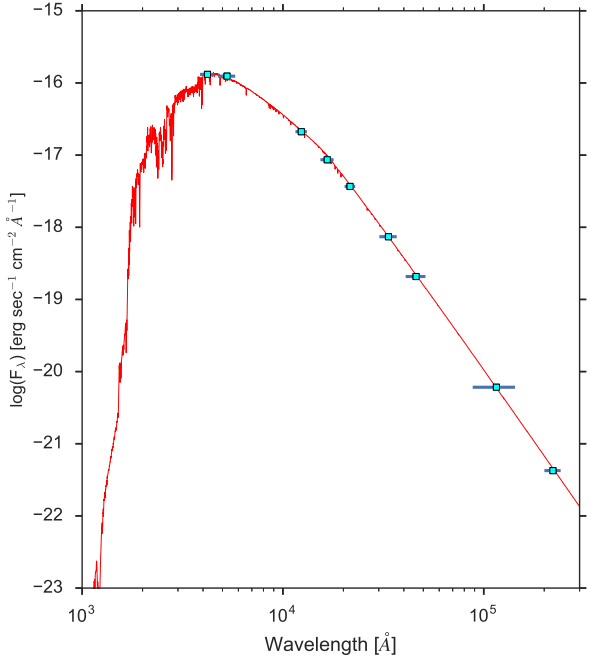}   
\caption{The spectral energy distribution of \thisstar with a model using the inferred stellar parameters from {\sc MINES}{\lowercase {\it weeper}}. The light blue points show the Tycho-2 B$_{T}$, V$_{T}$, 2MASS J, H, K$_{s}$, and WISE 1--4 photometric observations included in the fit. The red curve represents the best fit MIST stellar model.}
\label{SED_Fit}
\end{figure}

\subsection{TRES Spectroscopy}
\label{sec:Spec}
To measure the spectroscopic parameters of \thisstar, we used the Tillinghast Reflector Echelle Spectrograph (TRES) on the 1.5m telescope at the Fred L. Whipple Observatory (FLWO) on Mt. Hopkins, AZ. TRES is a fiber-fed echelle spectrograph, with a spectral resolving power of $\lambda / \Delta \lambda = 44000$. We observed \thisstar\ twice: on UT 2016 December 26 we obtained a 150 second exposure with a signal-to-noise ratio per resolution element of 41.7 at over the peak of the Mg b line order, and on UT 2017 January 8, we obtained a 990 second exposure, yielding a signal-to-noise ratio of 122. After cross-correlating the stellar spectra with synthetic templates, we find no evidence of a second set of spectral lines or any sign of a nearby blended source. The radial velocities measured from the two spectra differ by only 66 \ms, consistent with the expected RV uncertainties. After applying a correction to place the velocity of \thisstar\ on the IAU standard system, we measure an absolute RV of -3.65$\pm$0.1 \kms. From the stronger spectrum, we measured the star's Mt. Wilson activity indices $S_{\rm HK} = 0.166 \pm 0.004$ and $\log{R'_{\rm HK}} = -4.90 \pm 0.02$. Our measurements of the Mt. Wilson activity indicators were calibrated by comparing activity measuremenfs from TRES observations of stars also observed in the Mt. Wilson survey by \citep[][]{Duncan:1991}.  

Using the Stellar Parameter Classification tool (SPC), we inferred that \thisstar has a \teff = 6251$\pm$52 K, \mh = -0.27$\pm$0.08, log(g) = 4.1$\pm$0.1 (cgs), and a projected rotational velocity of 14.6$\pm$0.5 \kms\ \citep{Buchhave:2012, Buchhave:2014}. SPC determines these parameters by cross-correlating the observed stellar spectra with a grid of synthetic spectra from \citet{Kurucz:1992}. The synthetic spectra used by SPC includes a microturbulent velocity of 1.9 \kms. SPC does not model macroturbulence, so we also independently derived $v \sin I_\star$ and the macroturbulent velocity from least-squares deconvolution line profiles, derived from the TRES spectra \citep[following ]{Collier:2010}. We fit the least squares deconvolution broadening profiles simultaneously with the parameters $v\sin I_\star$ and the macroturbulent velocity, finding values of $v\sin I_\star = 12.9 \pm 0.4\,\mathrm{km\,s}^{-1}$, and v$_{mac}$ of $4.0 \pm 0.3\,\mathrm{km\,s}^{-1}$. This is consistent with the macroturbulence of late F-stars measured by \citet{Doyle:2014}, which were determined by spectroscopic follow-up of a series of \emph{Kepler} astro-seismic stars. We note that these errors are likely under-estimated since they do not include any systematic problems that may be present in the fitting or the LSD derivation.  

\begin{figure*}[!ht]
\vspace{0.3in}
\begin{tabular}{cc}
\includegraphics[width=0.5\linewidth]{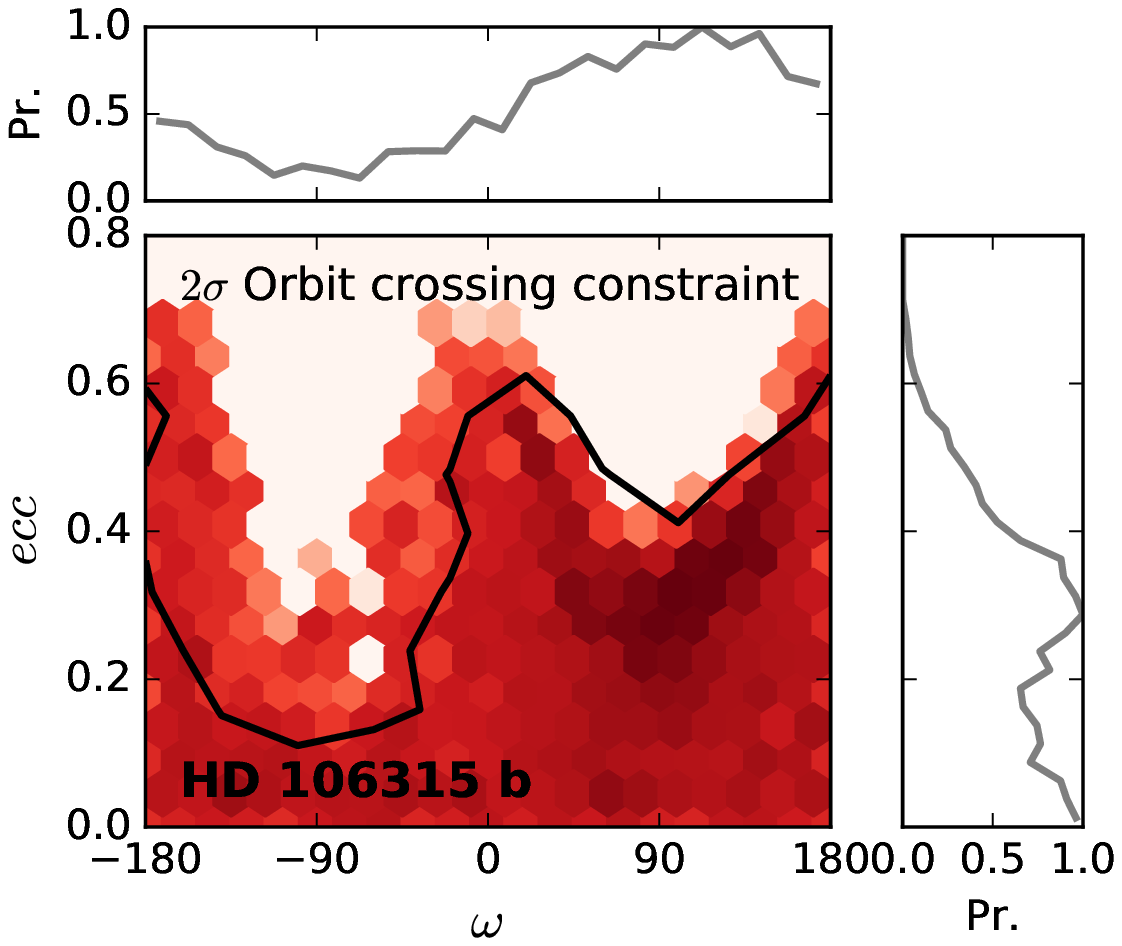} &
\includegraphics[width=0.5\linewidth]{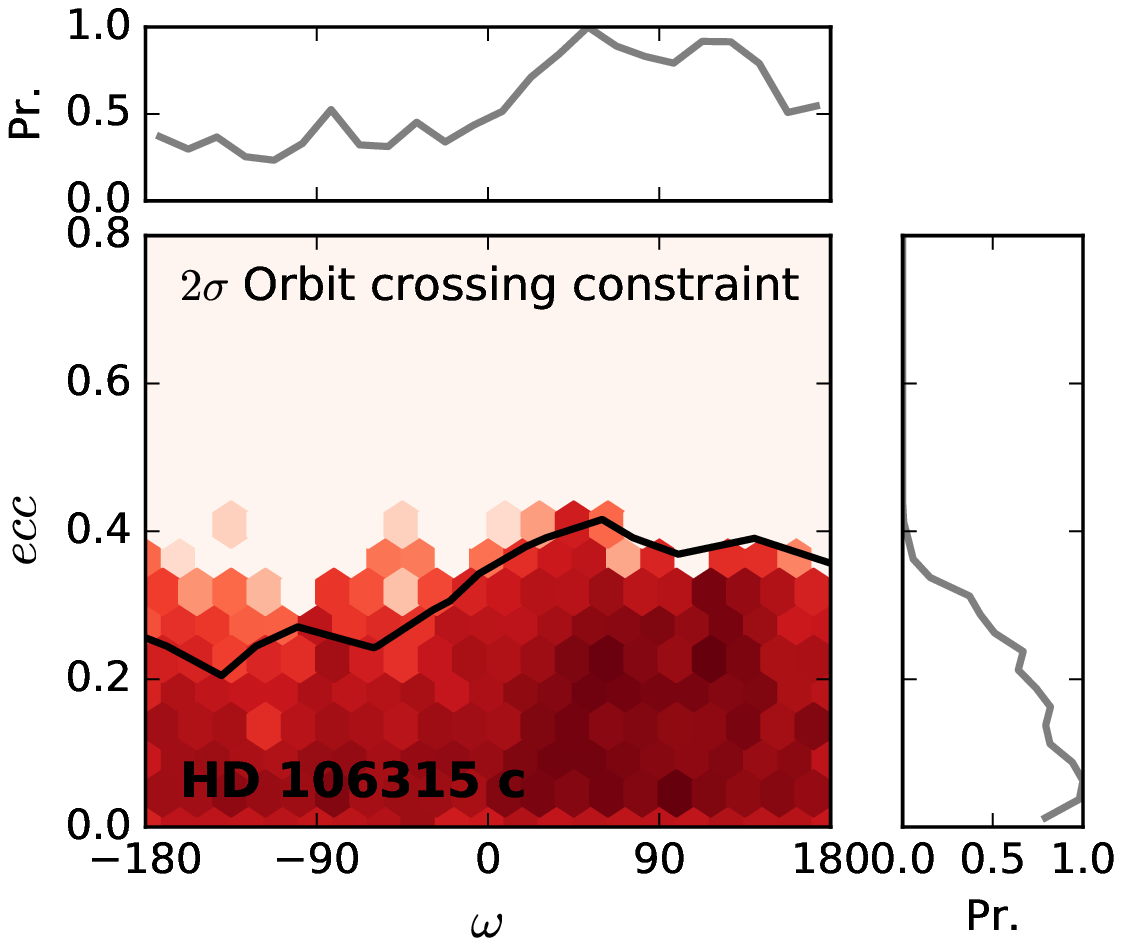}
\end{tabular}
\caption{Eccentricity $(ecc)$ and argument of periastron $(\omega)$ posterior probability distributions for \thisstar b (\textbf{left}) and c (\textbf{right}). The contours mark the regimes where 95\% of the solutions lie for the MCMC chains constrained by the orbit crossing criteria. In addition, the side panels mark the marginalized posteriors for eccentricity and $\omega$. }
\label{figure:ecc_posterior}
\end{figure*}

\subsection{Archival and Seeing-Limited Imaging}
To rule out the possibility of nearby bright companions, we visually inspected archival $J$-band observations of \thisstar from the 3.8 meter United Kingdom Infrared Telescope (UKIRT) located on Mauna Kea. We also observed \thisstar with KeplerCam on the 1.2m telescope at FLWO. KeplerCam has a 23$\arcmin$ $\times$ 23$\arcmin$ field-of-view and is binned 2$\times$2 resulting in a 0.67$\arcsec$ pixel scale. Additionally, we observed \thisstar with one of the eight MEarth-South telescopes. MEarth-South is located at the Cerro Tololo Inter-American Observatory in Chile and consists of eight 0.4m telescopes each with a 29$\arcmin$ $\times$ 29$\arcmin$ field-of-view and a  0.85$\arcsec$ pixel scale. From the combined archival and seeing-limited images, we confidently rule out any bright companions to \thisstar\ outside of about an arcsecond, and rule out any other stars inside the K2 photometric aperture at larger distances.



\section{System Modeling} 
\subsection{Spectral Energy Distribution and Stellar Properties}
\label{sec:SED}
To determine the stellar properties of \thisstar, we model all available photometry using {\sc MINES}{\lowercase {\it weeper}} (Cargile et al. in prep). {\sc MINES}{\lowercase {\it weeper}} is newly developed Bayesian approach for determining stellar parameters using the newest MIST stellar evolution models \citep{Choi:2016}. A detailed description of {\sc MINES}{\lowercase {\it weeper}} is given in Cargile et al. (in prep) and a brief summary can be seen in \S4 of \citet{Rodriguez:2017}. Unlike the case of V1334 Tau \citep{Rodriguez:2017}, we use the SPC spectral results (\feh\footnote{Note: While SPC measures metallicity (\mh), (that is, holding abundance ratios fixed at solar values and only adjusting the overall metal content), throughout our analysis, we use \mh\ and \feh\ interchangably.}, \teff, and Log(g); See \S \ref{sec:Spec}) and Gaia TGAS parallax measurements as priors in our analysis \citep{Gaia:2016}. We excluded the APASS photometry from this analysis due to our past experience with unaccounted for zero-point offsets (Cargile et al. in prep). Our final SED model is shown in Figure \ref{SED_Fit}. The determined stellar parameters are: Stellar Age = 3.987$^{+0.802}_{-0.516}$ Gyr, $M_{\star}$ = 1.105$^{+0.028}_{-0.036}$ \msun, $R_{\star}$ = 1.286$^{+0.039}_{-0.040}$ \rsun, log($L_{\star}$) = 0.368$^{+0.028}_{-0.026}$  \lsun, \teff = 6300$\pm$37 K, log($g$) = 4.261$^{+0.027}_{-0.024}$ cgs, \feh$_{initial}$ = $-$0.128$^{+0.041}_{-0.065}$ (metallicity at formation), \feh$_{surface}$ = $-$0.268$^{+0.060}_{-0.071}$ (current stellar metallicity), Distance = 109$\pm$3 Pc, $Av$ = 0.005$^{+0.027}_{-0.001}$ mag. This analysis used the MIST stellar evolution models while our global analysis used the YY Isochrones or the Dartmouth stellar evolution models. We use this analysis to check the global model determined stellar parameters (\S \ref{sec:GlobalModel}). Additionally, the SED analysis shows no sign of any IR excess and is consistent with a \teff = 6300 K stellar photosphere.


\subsection{Global Model}
\label{sec:GlobalModel}
We make use of the flattened \emph{K2} light curves, GAIA parallax, and TRES spectroscopic stellar parameters to perform a global modeling of the \thisstar system. Two independent analyses are presented in Table~\ref{tab:sysparams}, incorporating the Dartmouth \citep{Dotter:2008} and YY \citep{Yi:2001} isochrones. The systematic errors between the different stellar models used in the global model and \S \ref{sec:SED} are the likely cause of the $\sim$1.5$\sigma$ discrepancy in age and $\sim$2$\sigma$ difference in stellar mass. We note that the stellar radius between all stellar models is consistent (see Table \ref{tab:sysparams}). 

The Dartmouth models are incorporated in a global analyses (labeled ``Dartmouth'' in Table \ref{tab:sysparams}), with the light curves modeled using a modified version of \emph{EBOP} \citep{Popper:1981,Nelson:1972,southworth:2004}. The models are determined by the transit parameters of each planet: transit centroids $T_0$, periods $P$, radius ratios $R_p/R_\star$, normalized orbital radii $a/R_\star$, orbit inclination $i$, and eccentricity parameters $\sqrt{e} \cos \omega$ and $\sqrt{e} \sin \omega$. The quadratic limb darkening coefficients are fixed to those interpolated as per \citet{Sing:2010}. In addition, we incorporate the effective temperature $T_\mathrm{eff}$ and metallicity ([M/H]) with tight Gaussian priors into our analysis. At each iteration, we calculate a stellar density $\rho_\star$ from the transit parameters as per \citet{Seager:2003,Sozzetti:2007}, and fit these stellar parameters to the Dartmouth isochrones \citep{Dotter:2008} to derive a distance modulus. The distance modulus is then compared with the GAIA parallax via a likelihood penalty, further constraining the stellar and transit parameters. 

The parameter space is explored via a Markov chain Monte Carlo (MCMC) exercise, using the \emph{emcee} affine invariant ensemble sampler \citep{ForemanMackey:2012}. To avoid dynamically unstable solutions, links of the MCMC chain where the two orbits cross the Hill sphere of \thisstar b are removed. To calculate the Hill radius, we estimate the mass of each planet via mass-radius relationships from \citet{Weiss:2014} $(R_p < 4\,R_\oplus)$ and \citet{Lissauer:2011} $(R_p > 4\,R_\oplus)$. The resulting Hill spheres of b and c are small (0.002 AU), such that this constraint is essentially the removal of orbital crossing solutions in the MCMC chain. We also restrict the solutions such that the surface of \thisstar b does not extend beyond its Roche lobe at periastron \citep[following ][]{Hartman:2011}. Since the radius of \thisstar b is small, this is a also weak constraint that only removes the highest eccentricity solutions for this system. The best fit results are presented in Table~\ref{tab:sysparams} and Figure \ref{figure:LC}, and the derived eccentricity posteriors for each planet are shown in Figure~\ref{figure:ecc_posterior}.

We also model the \thisstar system via the new EXOFASTv2 code (labeled ``YY'' in Table \ref{tab:sysparams}). EXOFASTv2 (Eastman et al., in prep) is based on EXOFAST \citep{Eastman:2013} but many of the high-level codes were re-written from the ground up to be more general and flexible, allowing us to fit multiple planets, multiple sources of radial velocities, multiple transits with non periodic transit times, different normalizations, or separate band passes with just command line options and configuration files. It is also far easier to add additional effects and parameters.

EXOFASTv2 retains the global consistency of the model star and planet that the previous version had, but with the ability to fit multiple planets, and we now apply additional global constraints, like requiring that each planet's transit model implies the same stellar density \citep{Seager:2003} and their orbits are stable (i.e., they do not cross into other planets' Hill Spheres). Similar to the way EXOFAST handles limb darkening coefficients, EXOFASTv2 imposes a Gaussian prior derived from the \citet{Claret:2011} quadratic limb darkening tables given each step's value for the log($g$), \teff, and \feh, assuming model uncertainties of 0.05 in each limb darkening parameter.

The major conceptual departures from the original EXOFAST are that we use Yonsei-Yale isochrones \citep{Yi:2001} to simultaneously model the star instead of the Torres relations \citep{Torres:2010} (see, \citet{Eastman:2016} for a more detailed description), we replace the stepping parameter \logg \ with age, for which uniform priors should be more physical, and we replace the stepping parameter $\log(\ar)$ with $\log(\mstar)$ which allows for a more straight-forward extension to multiple planets since the semi-major axis is trivially derived from \mstar \ and the period using Kepler's law. We also now fit an added variance term instead of scaling and fixing the uncertainties.



The two major conceptual departures from the independent analysis above is that we use the YY stellar models instead of Dartmouth and we use the \citet{Chen:2017} exoplanet mass-radius relation (instead of \citet{Weiss:2014} and \citet{Lissauer:2011}) to estimate the planetary masses in order to exclude planetary eccentricities that would drive them into each other's Hill spheres.

We have used this system to validate EXOFASTv2 in addition to fully characterize the system. We include spectroscopic priors from the SPC analysis on \logg, \teff, and \feh. The results of the EXOFASTv2 fit can be seen in Table \ref{tab:sysparams}. All determined values for the YY and Dartmouth separate global fits are consistent with each other to $\sim$1$\sigma$. We present both the Dartmouth and YY results in Table \ref{tab:sysparams}. While we have no reason to prefer one over the other, for concreteness, we adopt the Dartmouth global model results for our discussion. The eccentricity of both planets is consistent with circular, with limits of $<$0.30 for planet b and $<$0.066 for planet c at 95\% confidence.


\begin{figure}
\begin{flushleft}
\includegraphics[width=0.99\linewidth, angle =-90, trim = 0 1in 0 0]{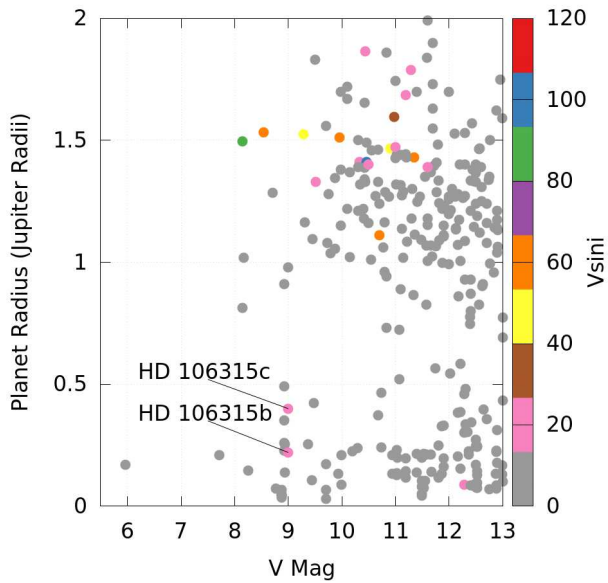}
\caption{The measured planet radius for all confirmed transiting planets brighter than $V$ = 13.0 color-coded by $v\sin I_*$. This figure was created using Filtergraph \citep{Burger:2013}.}
\label{fig:Vmag}
\end{flushleft}
\end{figure}

\begin{figure*}
\centering
\begin{tabular}{cc}
\includegraphics[width=0.45\linewidth]{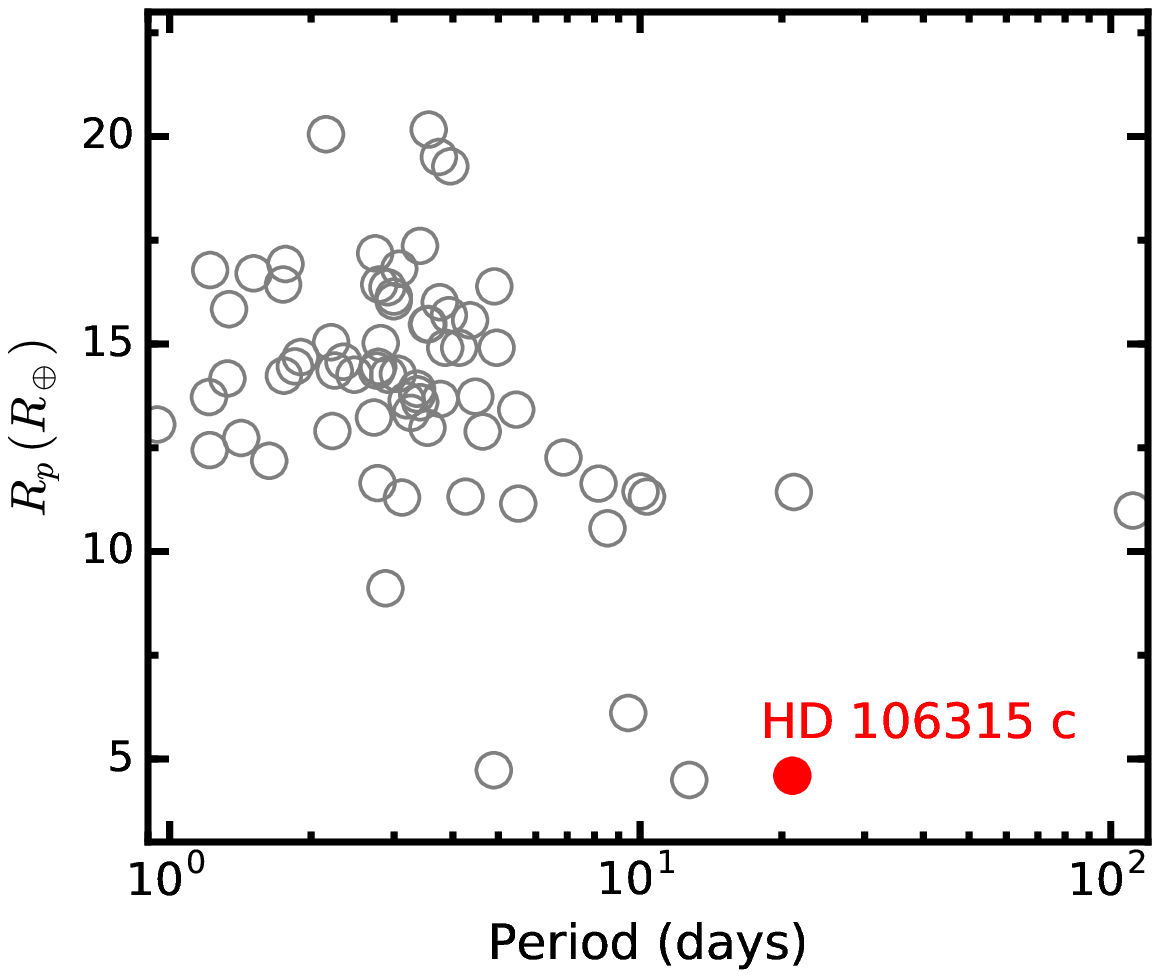} &
\includegraphics[width=0.5\linewidth]{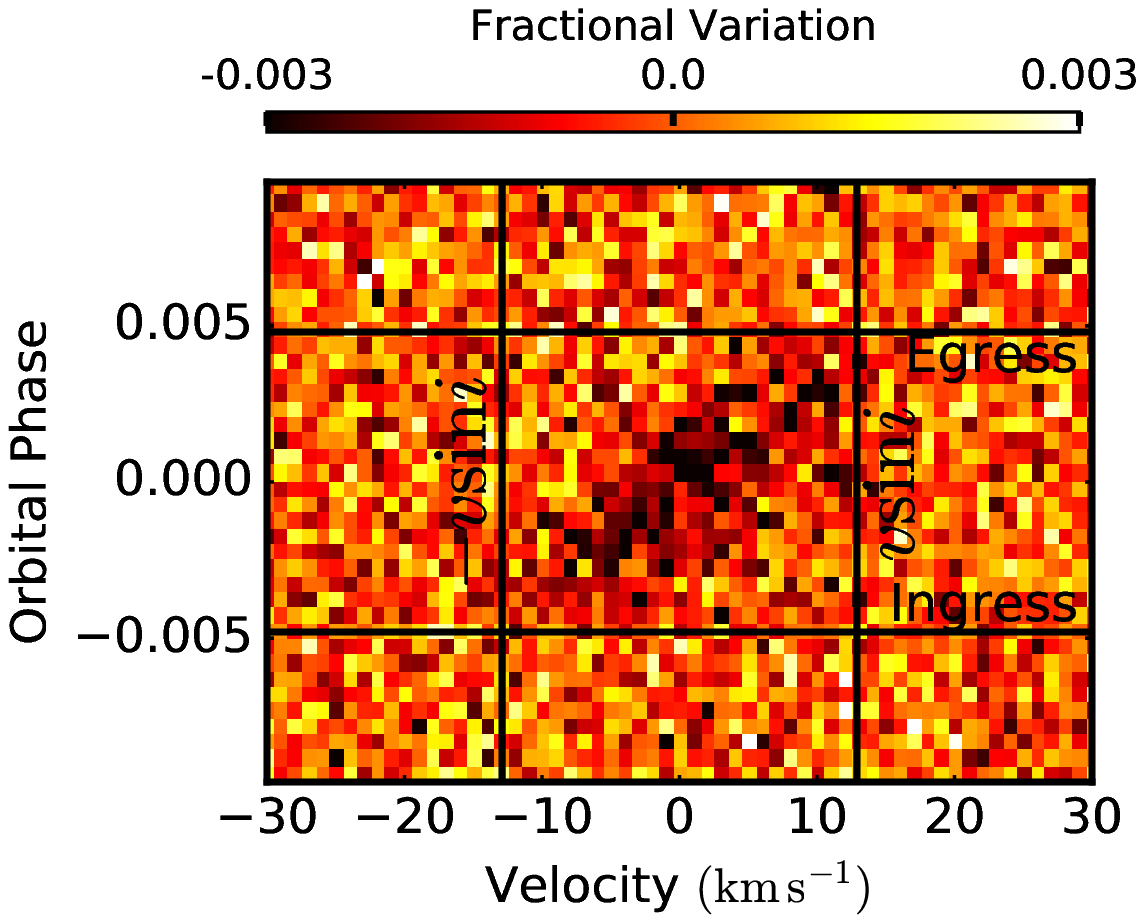}
\end{tabular}
\caption{\textbf{Left} The period -- planet radius distribution for all planets with secure spectroscopic obliquities measured. \thisstar c would be one of the longest period, smallest planets for which an obliquity can be measured. \textbf{Right} A simulated Doppler tomographic detection signal from observing a single transit of  \thisstar\ c using the MIKE. The ingress and egress are shown by the horizontal black lines and the boundaries of $-v\sin I_*$ and $v\sin I_*$ are marked by the vertical back lines.}
\label{fig:dt_simulation}
\end{figure*}

\begin{table*}
\footnotesize
\centering
\caption{\thisstar\ System parameters}
\begin{tabular}{llrr}
  \hline
  \hline
Parameter & Description & Dartmouth & YY \\
&& \bf{Adopted Value} &  Value\\
\hline
\emph{Stellar parameters} &&&\\
$M_\star\,(M_\odot)$       & Stellar mass          & $1.027_{-0.029}^{+0.034}$  & $1.086^{+0.042}_{-0.044}$ \\
$R_\star\,(R_\odot)$       & Stellar radius        & $1.281_{-0.058}^{+0.051}$  & $1.308^{+0.054}_{-0.050}$ \\
$T_\mathrm{eff}$ (K) $^a$  & Effective temperature & $6254_{-51}^{+55}$         & $6248^{+48}_{-47}$ \\
$\log \, g_\star$          & Surface gravity       & $4.234_{-0.033}^{+0.035}$  & $4.240\pm0.036$ \\
$\mathrm{[M/H]}$ $^b$      & Metallicity           & $-0.278_{-0.073}^{+0.082}$ & $-0.279^{+0.076}_{-0.074}$ \\
Age (Gyr)                  & Age                   & $5.91_{-0.79}^{+0.90}$     & $4.68^{+1.1}_{-0.94}$ \\
$L_\star\,(L_\odot)$       & Luminosity            & $2.24_{-0.16}^{+0.18}$     & $2.34^{+0.18}_{-0.16}$ \\ 
Distance (pc)              & Distance              & $109.7_{-4.4}^{+4.2}$      & $107.5^{+3.9}_{-3.6}$ \\
&&&\\
\emph{Planet b} &&&\\
$P$ (d)            & Orbital period            & $9.55385_{-0.00072}^{+0.00095}$ & $9.55496^{+0.00091}_{-0.00096}$ \\
$T_0$ (\bjdtdb)    & Transit centroid timing   & $247615.2057\pm0.0017$          & $2457615.2063^{+0.0015}_{-0.0016}$ \\
$R_p\,(R_\oplus)$  & Planet radius             & $2.40\pm0.12$                   & $2.56^{+0.14}_{-0.13}$ \\
$R_p\,(R_J)$       & Planet radius             & $0.214\pm0.011$                  & $0.228^{+0.013}_{-0.012}$ \\
$R_p/R_\star$      & Radius ratio              & $0.01717_{-0.00055}^{+0.00069}$ & $0.01792^{+0.00054}_{-0.00053}$ \\
$a/R_\star$        & Normalized orbital radius & $14.86_{-0.52}^{+0.64}$         & $15.00^{+0.60}_{-0.58}$ \\
$i\,(^\circ)$    & Orbit inclination       & $87.62_{-0.44}^{+1.59}$         & $87.61^{+0.97}_{-0.34}$ \\
$b$                & Impact parameter          & $0.63_{-0.43}^{+0.08}$          & $0.59^{+0.11}_{-0.31}$ \\
$a$ (AU)           & Orbital distance          & $0.08887_{-0.00093}^{+0.00082}$    & $0.0912\pm0.0012$ \\
$e$                & Eccentricity              & $<0.31\,(1\sigma)$ $^c$         & $0.25\,(1\sigma)$ $^c$ \\
$\omega\,(^\circ)$ & Argument of periastron    & $67_{-134}^{+72}$               &  $89^{+84}_{-85}$ \\
$T_{eq}$ (K)       & Equilibrium temperature   & $1146_{-22}^{+19}$              & $1140\pm20$ \\
$T_{14}$ (days)    & Total duration            & $0.159_{-0.009}^{+0.041}$       & $0.1548^{+0.0036}_{-0.0035}$ \\
$\tau$ (days)      & Ingress/egress duration   & $0.00444_{-0.00089}^{+0.00082}$ & $0.0042\pm0.0013$ \\
$T_S$ (\bjdtdb)    & Time of occultation       & $2457610.6\pm1.1$               &  $2457619.99^{+0.94}_{-0.93}$\\
&&&\\
\emph{Planet c} &&&\\
$P$ (d)            & Orbital period            & $21.0580_{-0.0022}^{+0.0022}$      & $21.0575\pm0.0014$ \\
$T_0$ (\bjdtdb)    & Transit centroid timing   & $2457611.1328_{-0.0014}^{+0.0015}$ & $2457611.13263^{+0.00097}_{-0.00099}$ \\
$R_p\,(R_\oplus)$  & Planet radius             & $4.40_{-0.27}^{+0.25}$             & $4.50^{+0.24}_{-0.22}$ \\
$R_p\,(R_J)$       & Planet radius             & $0.393_{-0.024}^{+0.022}$          & $0.401^{+0.021}_{-0.020}$\\
$R_p/R_\star$      & Radius ratio              & $0.03207_{-0.0011}^{+0.0009}$    & $0.03159^{+0.00075}_{-0.00085}$ \\
$a/R_\star$        & Normalized orbital radius & $25.69_{-1.1}^{+1.2}$               & $25.70^{+1.0}_{-0.98}$ \\
$inc\,(^\circ)$    & Transit inclination       & $88.48_{-0.18}^{+0.19}$            & $88.51^{+0.39}_{-0.18}$ \\
$b$                & Impact parameter          & $0.688_{-0.094}^{+0.044}$          & $0.61^{+0.11}_{-0.23}$ \\
$a$ (AU)           & Orbital distance          & $0.1503_{-0.0015}^{+0.0015}$       & $0.1564^{+0.0021}_{-0.0022}$ \\
$e$                & Eccentricity              & $<0.18\,(1\sigma)$ $^c$           &  $<0.23\,(1\sigma)$ $^c$\\ 
$\omega\,(^\circ)$ & Argument of periastron    & $53_{-134}^{+77}$                 & $89\pm61$ \\
$T_{eq}$ (K)       & Equilibrium temperature   & $874_{-19}^{+18}$                  & $871\pm15$ \\
$T_{14}$ (days)    & Total duration (days)     & $0.1970_{-0.0027}^{+0.0035}$       & $0.1939^{+0.0029}_{-0.0028}$ \\
$\tau$ (days)      & Ingress/egress duration   & $0.0115\pm0.0017$                 & $0.0092^{+0.0028}_{-0.0026}$ \\
$T_S$ (\bjdtdb)    & Time of eclipse           & $2457600.55_{-0.69}^{+0.49}$       & $2457600.6\pm1.6$\\

\hline
\hline
\end{tabular}
\begin{flushleft} 
\footnotesize{ \textbf{\textsc{NOTES:}}
$^a$ $T_\mathrm{eff}$ is constrained by a Gaussian prior about its spectroscopically determined parameters\\
$^b$ [M/H] is constrained by a Gaussian prior about its spectroscopically determined parameters\\
$^c$ Solutions with Hill sphere crossings have been removed.\\
}
\end{flushleft}
\label{tab:sysparams}
\end{table*}

\section{Statistical Validation}

Occasionally, transit signals like those we see in the K2 light curve of \thisstar\ can be caused by astrophysical phenomena other than transiting planets. We calculated the probability of such scenarios for the \thisstar\ planet candidates using \texttt{vespa} \citep{morton2015}, an implementation of the statistical procedure described by \citet{morton2012} to determine the likelihood that a transit signal is caused by a {\em bona fide} exoplanet. \texttt{Vespa} calculates false positive probablities for transiting planet candidates using information about the shape of the transits, constraints on the presence of other nearby stars, the location of the star in the sky (and hence the likelihood of an undetected background star contaminating the light curve), constraints in the difference in transit depth between even and odd transits, and constraints on the depth of putative secondary eclipses. \texttt{Vespa} considers several false positive scenarios, including the possibility that the transit signals are due to foreground eclipsing binary stars. For \thisstar, we include the additional constraint since we rule out an eclipsing binary scenario by our two different TRES observations showing no significant radial velocity difference. 

With \texttt{vespa}, we calculate false positive probabilities of $5\times10^{-3}$ and $3\times10^{-6}$ for \thisstar\ b and c, respectively. Because \thisstar\ hosts multiple transiting planet candidates, it is even less likely the candidates transiting \thisstar\ are false positives. \citet{lissauer:2012} calculated a ``multiplicity boost,'' or decrease in false positive probablity for multi-transiting systems in the \Kepler\ field of about a factor of 25. Subsequently, \citet{Sinukoff:2016} and \citet{Vanderburg:2016} have calculated that the multiplicity boost for K2 candidates is similar in magnitude. Applying the multiplicity boost to the \thisstar\ planets decreases the false positive probabilities to $2\times10^{-4}$ and $10^{-7}$ for \thisstar\ b and c, respectively. We therefore conclude that \thisstar\ b and \thisstar\ c are validated as genuine exoplanets.  

\section{Discussion}
The \thisstar\ system stands out among the currently known population of transiting planets for several reasons. First, \thisstar\ is one of the brightest host stars to host sub-Jovian planets. We accessed the NASA Exoplanet Archive \citep{Akeson:2013} on 11 Jan 2017 and found only 8 stars hosting planets smaller than 0.5 \rj\ brighter than \thisstar. Out of these bright sub-Jovian hosts, \thisstar\ is the only star above the Kraft break with a rotational velocity greater than 10 \kms. While ground-based transit surveys have found giant planets around hot, rapidly rotating bright stars \citep{CollierCameron:2010,Hartman:2015,Zhou:2016}, the \thisstar\ system is the first example of a multi-transiting system of small planets orbiting this type of star. Figure \ref{fig:Vmag} shows the \thisstar\ system in the context of other transiting planet systems. The brightness, small planet radii, and fast rotation make the \thisstar\ planets attractive targets for follow-up observations.


\begin{table}
 \centering
 \caption{The Best Confirmed Planets for Transmission Spectroscopy with R$_P$ $<$ 5\rearth}
 \label{tbl:S/N}
 \begin{tabular}{cccc}
    \hline
    \hline
    Planet & R$_P$(\rearth) & S/N$^a$  & Reference \\
    \hline
 GJ 1214 b  & 2.85$\pm$0.20 & 1.00 & \citet{Charbonneau:2009}\\
 GJ 436 b  & 4.1697408 & 0.68 & \citet{Gillon:2007}\\
 GJ 3470 b  & 3.88$\pm$0.33 & 0.48 & \citet{Bonfils:2012}\\
 HAT-P-11 b  & 4.73$\pm$0.26 & 0.46 & \citet{Bakos:2010}\\
 55 Cnc e  & 1.91$\pm$0.08 & 0.41 & \citet{Dawson:2010}\\
 HD 97658 b  & 2.34$^{+0.17}_{-0.15}$ & 0.30 & \citet{Dragomir:2013}\\
 HD 3167 c & 2.85$^{+0.24}_{-0.15}$ & 0.26 & \citet{Vanderburg:2016c}\\
 \textbf{HD 106315 c} & $4.3_{-0.3}^{+0.2}$ & 0.22 &  {\em This work} \\
 K2-25 b  & 3.43$^{+0.95}_{-0.31}$ & 0.22 & \citet{Mann:2016}\\
 HIP 41378 d  & 3.96$\pm$0.59 & 0.19 &\citet{Vanderburg:2016b} \\
 HIP 41378 b  & 2.90$\pm$0.44 & 0.14 &\citet{Vanderburg:2016b} \\
 K2-32 d  & 3.76$\pm$0.40 & 0.13 & \citet{Dai:2016}\\
 K2-19 c  & 4.86$^{+0.62}_{-0.44}$ & 0.12 & \citet{Armstrong:2015} \\
 K2-28 b  & 2.32$\pm$0.24 & 0.12 &\citet{Hirano:2016} \\
 K2-32 c  & 3.48$^{+0.98}_{-0.42}$ & 0.12 & \citet{Dai:2016}\\
 Kepler-105 b  & 4.81$\pm$1.5 & 0.11 & \citet{Wang:2014}\\
 Kepler-411 c  & 3.27$^{+0.12}_{-0.067}$ & 0.11 & \citet{Morton:2016}\\
 \textbf{HD 106315 b} & 2.5$\pm$0.1 & 0.10 & {\em This work} \\
   \hline
    \hline
 \end{tabular}
\begin{flushleft} 
 \footnotesize{ \textbf{\textsc{NOTES:}}
$^a$The predicted signal-to-noise ratios relative to GJ 1214 b. 
}
\end{flushleft}
\end{table}

\subsection{Prospects for Doppler Tomography}

The large numbers of compact Neptune / super-Earth systems discovered by the primary \emph{Kepler} mission have prompted the renaissance of in situ formation models for planets with gaseous envelopes \citep{Lee:2014,Batygin:2016,Boley:2016}. The formation and migrational history of planetary systems are embedded in their present-day orbital obliquities. However, few multi-planet systems offer the opportunity for us to characterize their orbital obliquities via unbiased techniques. The obliquities of multi-planetary systems and longer period warm Jupiters can be measured via star-spot crossings \citep[e.g.][]{Sanchis-Ojeda:2012, Dai:2017}, and via asteroseismology for planets around evolved stars \citep[e.g.][]{Huber:2013,Quinn:2015}.  Figure~\ref{fig:dt_simulation} (left) shows the set of planets for which obliquities have been measured spectroscopically\footnote{From the Holt-Rossiter-McLaughlin Encyclopedia \url{http://www2.mps.mpg.de/homes/heller/}}. WASP-47 is the only multi-transiting system with a spectroscopically measured obliquity \citep{Becker:2015,Sanchis-Ojeda:2015}.

The possibility of further obliquity characterization for the \thisstar system makes this discovery especially important. \thisstar is a V=9.0 star with a $v \sin I_*$ of $12.9\,\mathrm{km\,s}^{-1}$, making it an excellent target for further follow-up via Doppler tomography \citep[e.g.][]{Collier:2010,Johnson:2014,Zhou:2016}. Figure~\ref{fig:dt_simulation} (right) shows a simulated Doppler tomographic transit of \thisstar\,b, as observed by the Magellan Inamori Kyocera Echelle (MIKE) spectrograph on the 6.5m Magellan Clay telescope. We assume an exposure time of 15\,min, each with a signal-to-noise scaled from that of the TRES spectra presented in Section~\ref{sec:Spec}, and spectral resolution convolved to that of MIKE ($\lambda / \Delta \lambda = 65000$). A macroturbulence broadening of $4\,\mathrm{km\,s}^{-1}$ has also been included, accounting for additional broadening of the line profiles. The planetary Doppler tomography signal is detected at a significance of $7 \,\sigma$. 

\subsection{Prospects for Transmission Spectroscopy}
Transmission spectroscopy will also be a powerful diagnostic of the system's formation history. The planets' atmospheric compositions depend on their origin: for example, a planet forming outside the water ice line can accrete water-rich planetesimals, whereas we would expect closer-in formation locations to lead to a drier composition \citep{Chianglaughlin:2013}.

\thisstar stands out as a prime system for atmosphere characterization thanks to the brightness of the host star (H mag = 8). As shown in Table \ref{tbl:S/N}, both planets rank in the top twenty best small planets for transmission spectroscopy measurements (R$_P < 5\rearth$). The signal-to-noise (SNR) calculations were made with the same assumptions as in \cite{Vanderburg:2016c}.

Although they are among the highest SNR candidates known, the \thisstar planets still pose a challenge for transmission spectroscopy with current facilities. To assess prospects for observing the system with HST/WFC3, the current state-of-the-art instrument for atmosphere studies, we calculated a model spectrum for the outer planet with the ExoTransmit code \citep{Kempton:2016}. We assumed a 100x solar metallicity atmosphere and a surface gravity equal to 7 m\,s$^{-2}$ (M$_P$ = 15 \mearth). For this case, the amplitude of spectral features is just 70 ppm, which is within reach of an intensive multi-transit observing campaign with HST \citep[e.g.][]{Kreidberg:2014, Line:2016}. However, if the planet mass is larger, the atmosphere is more enhanced in metals, or aerosols are present, the amplitude of spectral features will decrease, and they may not be detectable until JWST launches. 

A further complication is that the planet masses are unknown and will be challenging to measure. \thisstar\ is a rapid rotator ($v \sin{I_*} = 12.9$ \kms), which broadens the lines and inhibits precise RV measurements. According to the NASA Exoplanet Archive (accessed 10 Jan 2017), for all planet-hosting stars with a $v\sin I_*$ $>$ 10 \kms, the smallest measured RV semi-amplitude is 33 \ms, whereas the expected semi-amplitude for \thisstar c is only 2-5 \ms. The absence of mass measurements is problematic for interpreting the transmission spectrum, because atmospheric metallicity and surface gravity are highly degenerate \citep{Batalha:2017}. Therefore, to take full advantage of the potential this system has for precise transmission spectroscopy, it will be necessary to explore alternative prospects for measuring the masses. Recently, progress has been made measuring the masses of small planets around moderately rotating stars using advanced statistical techniques and high-cadence observations, \citep[see, for example, ][]{lopezmorales:2016}, but measuring the masses of the \thisstar\ planets will be an even greater challenge. Transit timing variations (TTVs) could be an alternate way to measure planet masses, but \thisstar\ b and c do not orbit particularly close to any strong mean motion resonances ($P_C / P_B \simeq 2.2$), so any TTVs will be small.

\section{Conclusion}
We present the discovery of two transiting planets orbiting the bright F-star \thisstar. \thisstar b is a sub-Neptune size planet with a radius of $2.5\pm0.1$ \rearth and a $9.5539_{-0.0007}^{+0.0009}$ day orbit. \thisstar c is a warm super-Neptune size planet with a radius of $4.3_{-0.3}^{+0.2}$ \rearth and an orbital period of $21.058\pm0.002$  days. The large rotational velocity of \thisstar provides an attractive opportunity to measure the spin-orbit angle for a Neptune sized planet. This measurement may provide evidence to distinguish whether \thisstar c formed in situ or farther out in the protoplanetary disk and migrated to its current location. Future observations should attempt to measure the mass of each planet, a parameter important for proper interpretation of any transit spectroscopy, but such mass determinations will likely require capabilities beyond what is presently achievable with precise RV measurements or transit timing variations. 

{\bf Note added in review:} During the preparation of this paper, our team became aware of another paper reporting the discovery of a planetary system orbiting \thisstar \citep{Crossfield:2017}. The results from both papers are consistent to each other. No information about the analysis procedure or any results were shared between groups prior to the submission of both papers.

\acknowledgements
Work performed by J.E.R. was supported by the Harvard Future Faculty Leaders Postdoctoral fellowship. A.V. is supported by the NSF Graduate Research Fellowship, Grant No. DGE 1144152. D.W.L. acknowledges partial support from the from the TESS mission through a sub-award from the Massachusetts Institute of Technology to the Smithsonian Astrophysical Observatory. Work performed by P.A.C. was supported by NASA grant NNX13AI46G. This research has made use of NASA's Astrophysics Data System and the NASA Exoplanet Archive, which is operated by the California Institute of Technology, under contract with the National Aeronautics and Space Administration under the Exoplanet Exploration Program. This paper includes data collected by the \Kepler\ mission. Funding for the K2 mission is provided by the NASA Science Mission directorate. Some of the data presented in this paper were obtained from the Mikulski Archive for Space Telescopes (MAST). STScI is operated by the Association of Universities for Research in Astronomy, Inc., under NASA contract NAS5--26555. Support for MAST for non--HST data is provided by the NASA Office of Space Science via grant NNX13AC07G and by other grants and contracts. UKIRT is supported by NASA and operated under an agreement among the University of Hawaii, the University of Arizona, and Lockheed Martin Advanced Technology Center; operations are enabled through the cooperation of the East Asian Observatory.

\bibliographystyle{apj}

\bibliography{HD106315}

\end{document}